\documentclass[sn-mathphys,Numbered]{sn-jnl}% Math and Physical Sciences Reference Style
\usepackage{graphicx}%
\usepackage{multirow}%
\usepackage{amsmath,amssymb,amsfonts}%
\usepackage{amsthm}%
\usepackage{mathrsfs}%
\usepackage[title]{appendix}%
\usepackage{xcolor}%
\usepackage{textcomp}%
\usepackage{manyfoot}%
\usepackage{booktabs}%
\usepackage{algorithm}%
\usepackage{algorithmicx}%
\usepackage{algpseudocode}%
\usepackage{listings}%
\usepackage{caption}%
\theoremstyle{thmstyleone}%
\theoremstyle{thmstyletwo}%

\theoremstyle{thmstylethree}%

\begin{document}

\title[Article Title]{The Future of ChatGPT-enabled Labor Market: A Preliminary Study in China}

%%=============================================================%%
%% Prefix	-> \pfx{Dr}
%% GivenName	-> \fnm{Joergen W.}
%% Particle	-> \spfx{van der} -> surname prefix
%% FamilyName	-> \sur{Ploeg}
%% Suffix	-> \sfx{IV}
%% NatureName	-> \tanm{Poet Laureate} -> Title after name
%% Degrees	-> \dgr{MSc, PhD}
%% \author*[1,2]{\pfx{Dr} \fnm{Joergen W.} \spfx{van der} \sur{Ploeg} \sfx{IV} \tanm{Poet Laureate} 
%%                 \dgr{MSc, PhD}}\email{iauthor@gmail.com}
%%=============================================================%%

\author[1]{\fnm{Lan} \sur{Chen}}
%\equalcont{These authors contributed equally to this work.}

\author[1]{\fnm{Xi} \sur{Chen}}
%\equalcont{These authors contributed equally to this work.}

\author[1]{\fnm{Shiyu} \sur{Wu}}
%\equalcont{These authors contributed equally to this work.}

\author[1]{\fnm{Yaqi} \sur{Yang}}
%\equalcont{These authors contributed equally to this work.}

\author[1]{\fnm{Meng} \sur{Chang}}

\author*[*]{\fnm{Hengshu} \sur{Zhu}}\email{zhuhengshu@kanzhun.com}
\affil[1]{\orgdiv{Career Science Lab}, \orgname{BOSS Zhipin}}

%%==================================%%
%% sample for unstructured abstract %%
%%==================================%%

\abstract{
As a phenomenal large language model, ChatGPT has achieved unparalleled success in various real-world tasks and increasingly plays an important role in our daily lives and work. However, extensive concerns are also raised about the potential ethical issues, especially about whether ChatGPT-like artificial general intelligence (AGI) will replace human jobs. To this end, in this paper, we introduce a preliminary data-driven study on the future of ChatGPT-enabled labor market from the view of Human-AI Symbiosis instead of Human-AI Confrontation. To be specific, we first conduct an in-depth analysis of large-scale job posting data in BOSS Zhipin, the largest online recruitment platform in China. The results indicate that about 28\% of occupations in the current labor market require ChatGPT-related skills. Furthermore, based on a large-scale occupation-centered knowledge graph, we develop a semantic information enhanced collaborative filtering algorithm to predict the future occupation-skill relations in the labor market. As a result, we find that additional 45\% occupations in the future will require ChatGPT-related skills. In particular, industries related to technology, products, and operations are expected to have higher proficiency requirements for ChatGPT-related skills, while the manufacturing, services, education, and health science related industries will have lower requirements for ChatGPT-related skills.}


%%\pacs[JEL Classification]{D8, H51}

%%\pacs[MSC Classification]{35A01, 65L10, 65L12, 65L20, 65L70}

\maketitle

\section{Introduction}\label{sec1}

Artificial intelligence (AI) is having an increasingly pervasive impact on our lives, and yet, its potential impact on our jobs has not been given enough attention. With the unparalleled success achieved by ChatGPT, a phenomenal large language model, more and more people are beginning to realize the extent of AI's involvement in our work. The ability of ChatGPT to understand and respond to people's instructions properly has amazed many. With its vast knowledge base and ability to process large amounts of data, ChatGPT is often able to provide better answers to questions than humans. Additionally, AI-generated images and music are now virtually indistinguishable from those created by humans. This could have a significant impact on jobs related to art and design, as AI can produce images far more efficiently than humans.

As awareness of the power of AI grows, extensive concerns are also raised about the potential ethical issues, especially about whether ChatGPT-like artificial general intelligence (AGI) will replace human jobs. In a recent paper published by OpenAI\cite{eloundou2023gpts}, it was stated that ChatGPT has the potential to impact 80\% of jobs in the United States, with around 19\% of workers seeing their tasks impacted by more than 50\%. The study indicates that occupations requiring writing and programming skills are more vulnerable to being replaced by language models, and higher-paying jobs may be more exposed. However, these predictions do not specify a timeframe, making it difficult to determine which jobs are currently being displaced by AI in reality, and how other occupations may be gradually affected over time.

To this end, in this paper, we aim to introduce a preliminary data-driven study on the future of ChatGPT-enabled labor market from the view of Human-AI Symbiosis instead of Human-AI Confrontation. To be specific, we first conduct an in-depth analysis of large-scale job posting data in BOSS Zhipin, the largest online recruitment platform in China. Furthermore, based on a large-scale occupation-centered knowledge graph, we develop a semantic information enhanced collaborative filtering algorithm to predict the future occupation-skill relations in the labor market. This analysis provides valuable insights into how ChatGPT may shape the labor market and identifies the key skills that will be in demand in the future.

\section{Results}\label{sec2}
\subsection{The current status of labor market enabled by ChatGPT}\label{subsec1}

As a state-of-the-art language model designed for natural language processing, ChatGPT has the potential to simulate human thinking patterns, comprehend and generate human language, and help people accomplish tasks more efficiently and with fewer errors in the workplace. As a result, it is foreseeable that possessing ChatGPT-related skills will enhance job seekers' competitiveness in a wide range of industries. Awareness has already been heightened on both sides of the labor market.

%%boss搜索词 trends
\begin{figure}[!h]
    \centering
    \includegraphics[width=10cm]{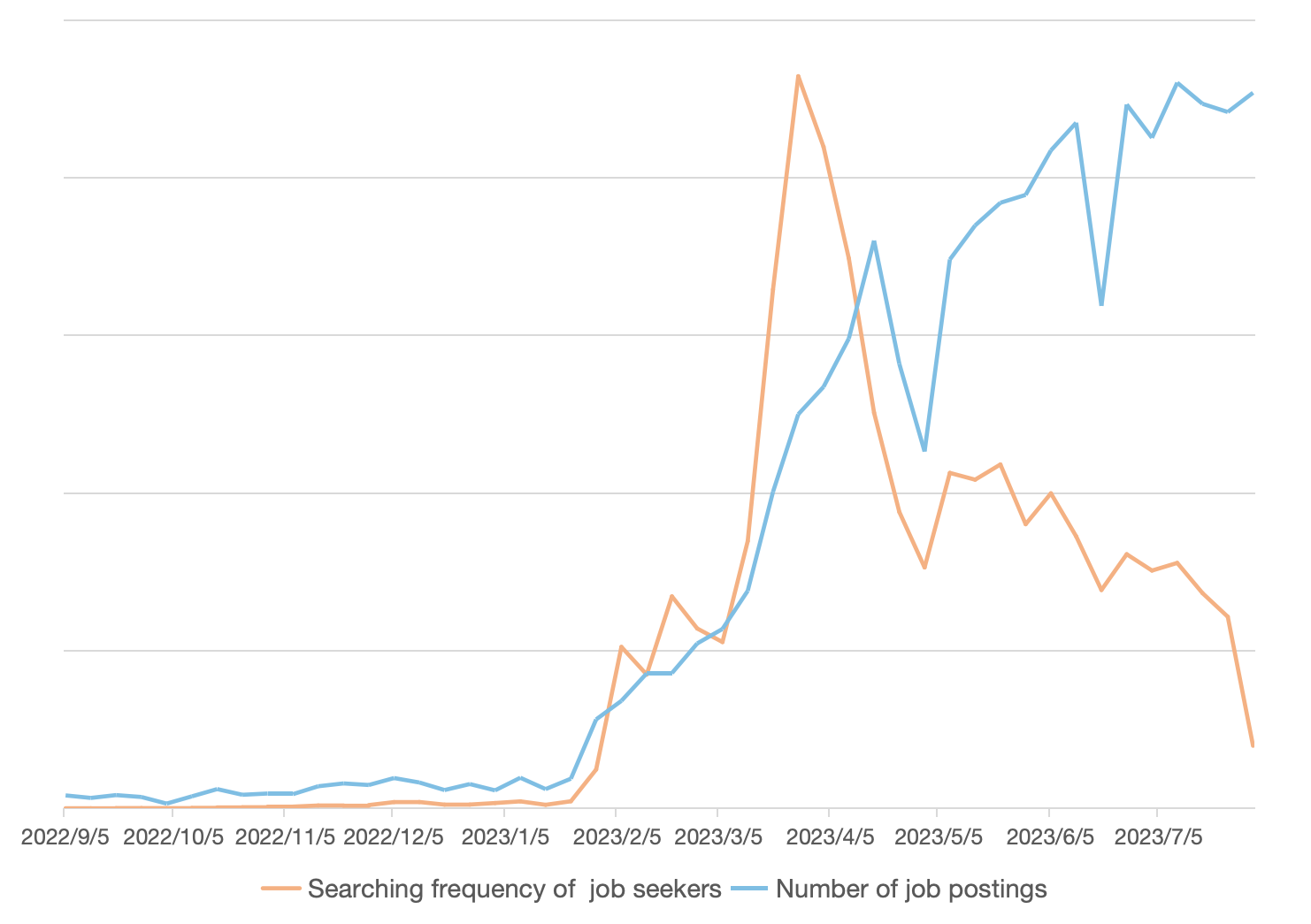}
    \caption{Comparison between the behaviors of job seekers and recruiters}
    \label{fig:1}
\end{figure}

The following graph illustrates the behavioral patterns of both job seekers and recruiters between September 5, 2022, and July 31, 2023, on BOSS Zhipin. It showcases the frequency with which job seekers searched for ChatGPT-related keywords and the concurrent release of job postings by recruiters. Given the recurrent nature of job postings, which typically exhibit minimal activity during weekends, we opted to aggregate this data on a weekly basis to provide a clearer perspective. It is crucial to emphasize that the two indices are inherently dissimilar in terms of quantity. Therefore, we undertook a standardization process to bring them to the same scale, rendering their trends directly comparable.

Evident from the graph, there was a notable shift in the perceptions of ChatGPT's importance among both job seekers and recruiters towards the end of January 2023, as indicated by a sudden upturn in both indices. However, considering the supply side of the labor market, the frequency of searches increased at a more rapid pace and reached its peak by the end of March. When contrasted with the search patterns for other recently emerged technologies, this trend stands out as a remarkable phenomenon. On the demand side, job postings exhibited a steady, uninterrupted rise, continuing until the conclusion of July, yet without reaching their peak. The two conspicuous declines in the posting volume can be attributed to two major national holidays, which led to reduced recruitment activity during those periods.

This paper will primarily focus on the demand side of the labor market, shedding light on its increasing appetite for skills related to ChatGPT. In this section, we will examine the industries and occupations that have begun to recruit employees with ChatGPT-related skills and their proficiency level requirements. Additionally, we will explore several new occupations that have emerged as a result of the advent of ChatGPT.

\subsubsection{Occupations that require ChatGPT-related skills}\label{subsubsec1}

According to our analysis of the BOSS Zhipin platform, there are currently 28\% occupations that require job seekers to have ChatGPT skills. As shown in Figure~\ref{fig:1}, these positions are mostly found in industries with higher salaries such as technology, operations, products, design, and sales. This aligns with OpenAI's prediction that higher-paying occupations are more likely to be affected by ChatGPT.

\begin{figure}[!h]
    \centering
    \includegraphics[width=10cm]{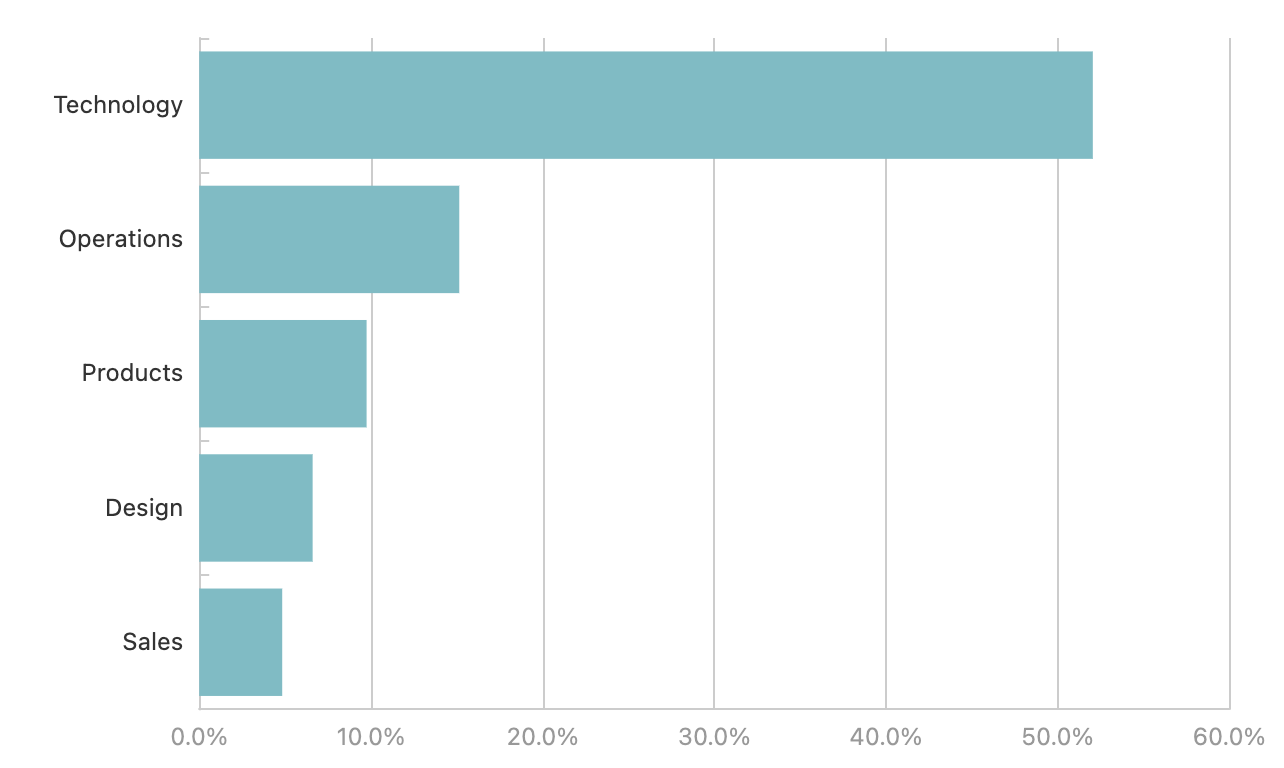}
    \caption{Distribution of occupations related to ChatGPT}
    \label{fig:1}
\end{figure}

Occupations related to computer technology are most closely related to ChatGPT, accounting for more than half of the postings. Most of these occupations are algorithm engineers, which are responsible for the development and implementation of various algorithm models. This occupation appears most frequently because some of them need to develop large language models(LLM) similar to ChatGPT. Algorithm engineers with ChatGPT-related skills are expected to have a strong understanding of natural language processing. This includes the ability to develop models that can understand, analyze, and generate human language, which is a critical component of many AI applications. Meanwhile, ChatGPT also helps algorithm engineers improve work efficiency and model quality.

Typical occupations in the operation sector include many occupations, such as online operations, community operations and content operations. This sector ranks second, which is understandable because many jobs related to operations require job seekers to have comprehensive abilities, sometimes including planning, market analysis, video editing, copywriting, etc. ChatGPT happens to be able to enhance all of these skills. It can help operations professionals generate various marketing solutions including advertisements, social media posts and emails; analyze large amounts of data to identify patterns and trends in business operations; automatically generate articles and provide text-editing suggestions.

Similarly, ChatGPT plays a significant role in product development, design, and sales sectors. In addition to the aforementioned aspects, its functions such as automated designing, video making, and natural language processing can help these industries improve work efficiency and customer satisfaction, causing it highly favored by many companies.

\subsubsection{Proficiency in ChatGPT-related skills}\label{subsubsec2}

Our investigation of ChatGPT proficiency requirements revealed that occupations in computer technology and product development sectors typically require a high level of proficiency, which is not surprising given their focus on LLM or AI product design and development. Media and marketing-related occupations, while fewer in number, also require a relatively high level of proficiency. Occupations such as advertising copywriters, activity planners, and marketing consultants require complex ChatGPT prompts to create efficient planning schemes, gain insight into social media platforms and user behavior, and respond better to market demand. On the other hand, simpler ChatGPT prompts suffice for translators, writers, and live stream operators who only need basic information from the model.

\subsubsection{ChatGPT related job postings offer higher salary}\label{subsubsec3}
Another intriguing yet unsurprising revelation we have uncovered is the propensity for job postings that include keywords associated with ChatGPT to offer salaries that surpass the occupation average. In Figure 2, we present a comparison between the average monthly salaries of specific occupations and those associated with ChatGPT-related roles within these occupational categories. The showcased occupations represent the top sectors that yield the highest number of job opportunities related to ChatGPT. To some extent, this supports our assertion that individuals who actively invest in learning how to leverage the capabilities of ChatGPT tend to gain a competitive advantage in the labor market.

\begin{figure}[!h]
    \centering
    \includegraphics[width=12cm]{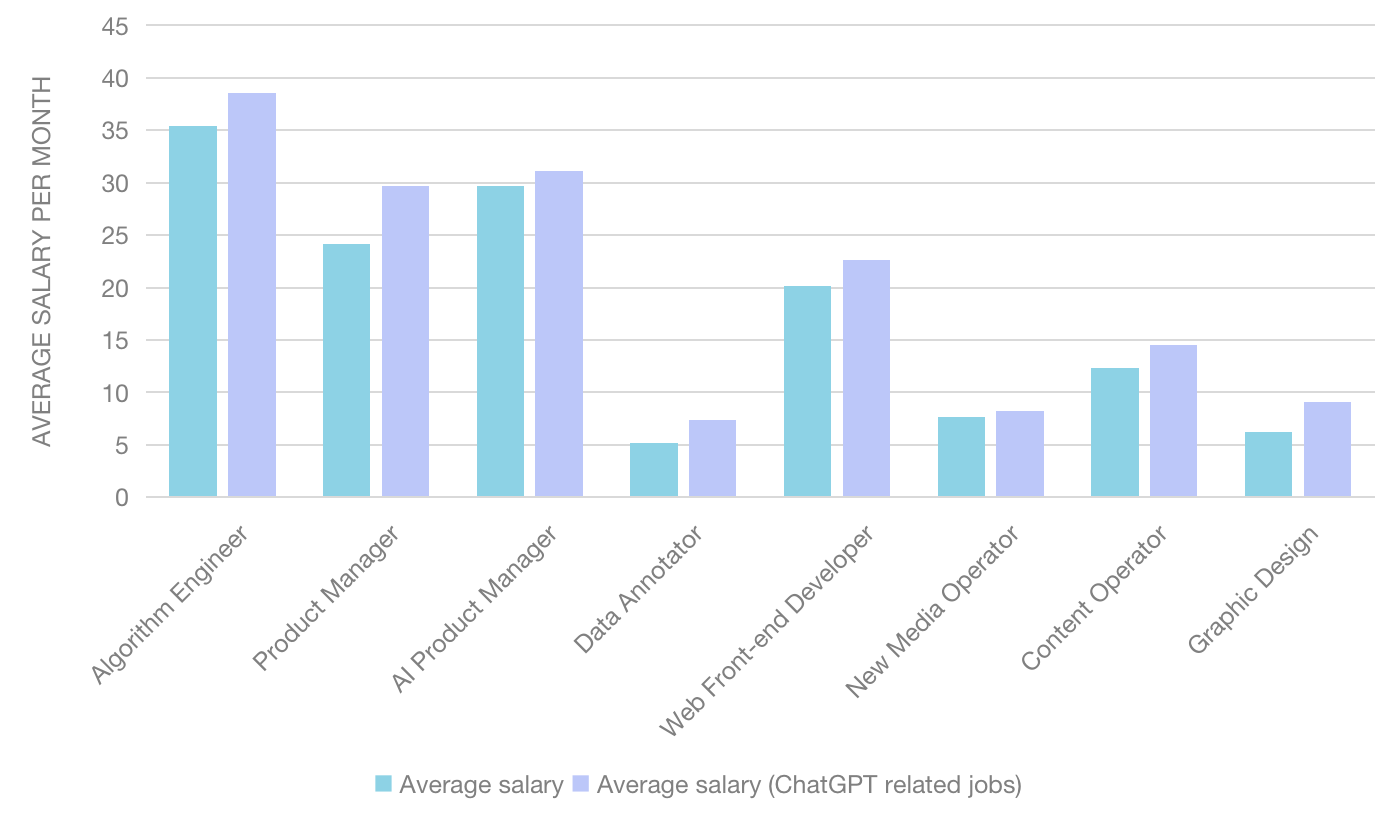}
    \caption{The average salary differences between ChatGPT related jobs and the occupation}
    \label{fig:wordcloud}
\end{figure}

\subsubsection{ChatGPT creates new job opportunities}\label{subsubsec3}

In addition to enabling existing professions, ChatGPT has also created new occupations or increased demand for certain roles, such as prompt word engineers, language model trainers, and ChatGPT optimization specialists. Prompt word engineers design appropriate prompts for ChatGPT to generate accurate code or content. Language model trainers fine-tune ChatGPT for specific conversation models by iterating training. ChatGPT optimization specialists identify areas where ChatGPT can optimize a company's existing business processes and enhance the model's performance, accuracy, and stability.

As the adoption of ChatGPT and other AI technologies continues to increase, job seekers must develop related skills to remain competitive in the labor market. In the future, having proficiency in ChatGPT and other AI technologies will become increasingly essential for workers across various industries, as these technologies continue to reshape the nature of work.

\subsection{Prediction of occupations enabled by ChatGPT}\label{subsec2}
\subsubsection{New occupations enabled by ChatGPT}\label{subsubsec1}

According to our predictions, ChatGPT is likely to enable additional 45\% occupations in the future (Figure~\ref{fig:wordcloud}). Of these occupations, around 7\% will require candidates to have advanced skills in using ChatGPT, while approximately one-third of the occupations will require basic knowledge of the technology and its applications. As shown in Figure~\ref{fig:2}, these new roles are expected to be in industries such as manufacturing, services, education, and technology. 

\begin{figure}[!h]
    \centering
    \includegraphics[width=10cm]{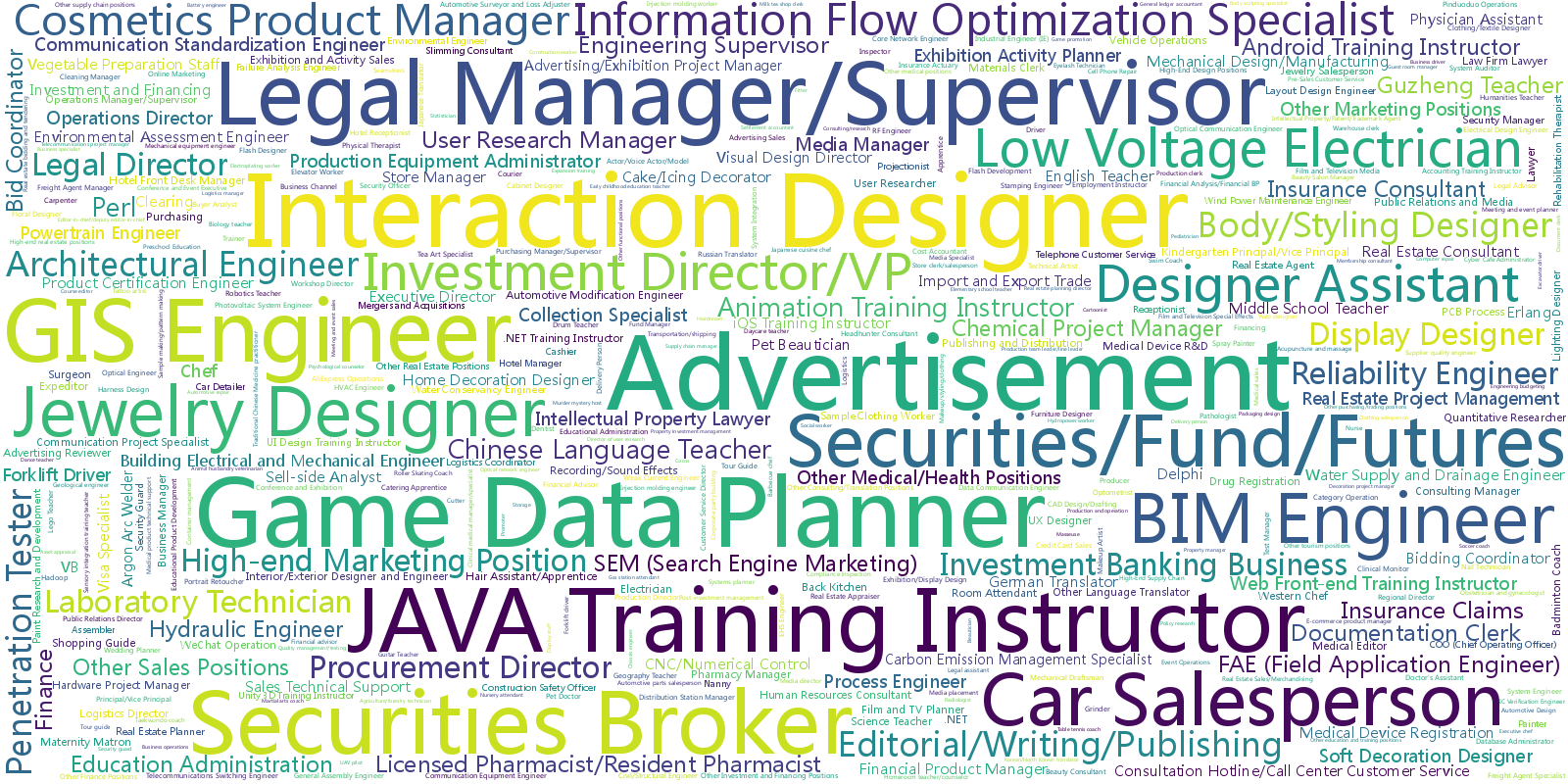}
    \caption{Word cloud of new occupations enabled by ChatGPT}
    \label{fig:wordcloud}
\end{figure}

\begin{figure}[!h]
    \centering
    \includegraphics[width=10cm]{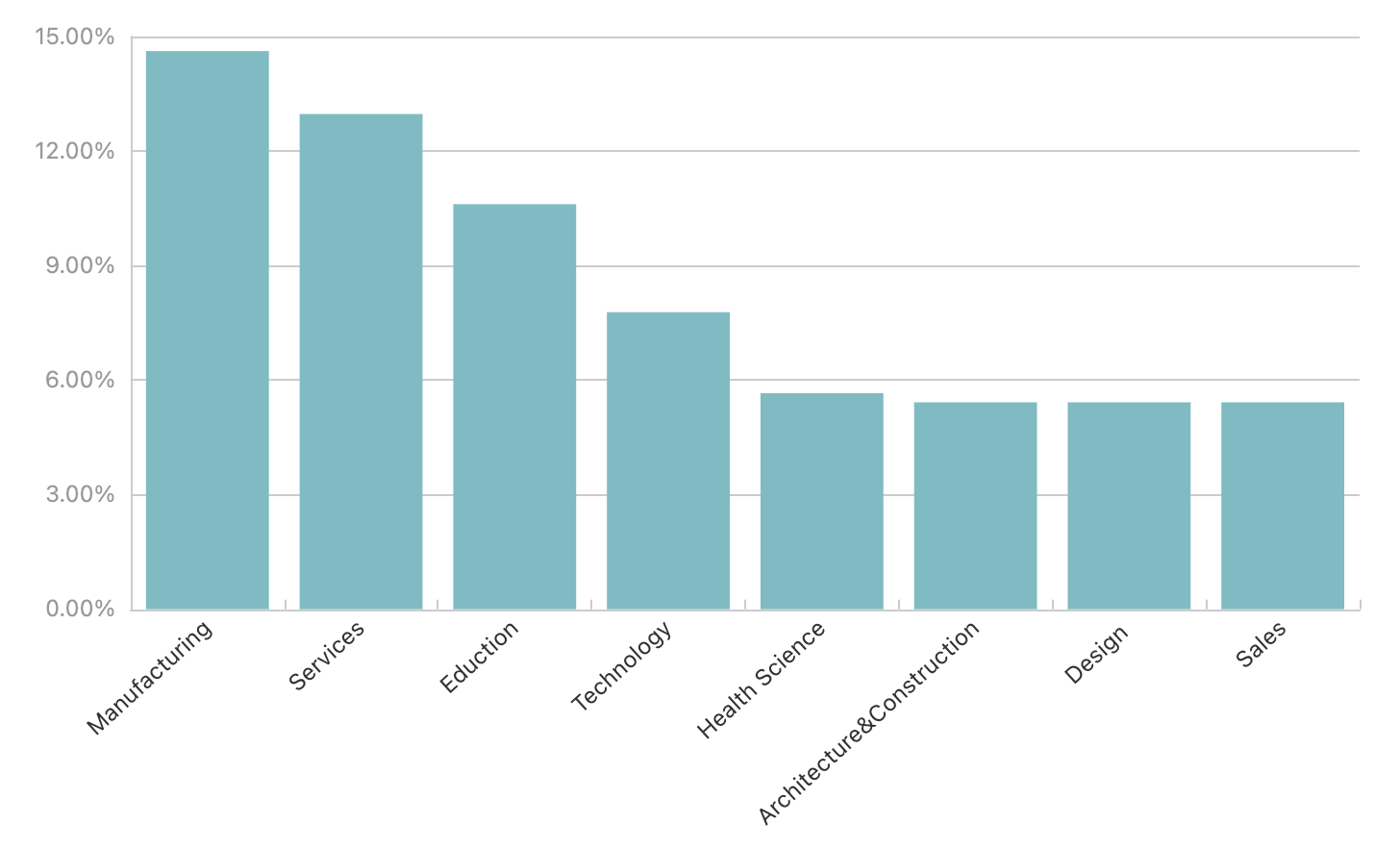}
    \caption{Distribution of new occupations enabled by ChatGPT}
    \label{fig:2}
\end{figure}

Although manufacturing, services, and education industries will have a higher number of ChatGPT-enabled occupations in the future, the proficiency requirements for job seekers will not be too high. Occupations such as mechanical drafters, sales promoters, and course editors may only require a basic understanding of AI tools like ChatGPT and their functions. However, technology-related occupations will have higher requirements for ChatGPT application skills, with candidates expected to have a deep understanding of ChatGPT, and even to be proficient in its use and principles. 

\subsubsection{Occupations with high proficiency requirements for ChatGPT-related skills}\label{subsubsec2}

\begin{figure}[!h]
    \centering
    \includegraphics[width=10cm]{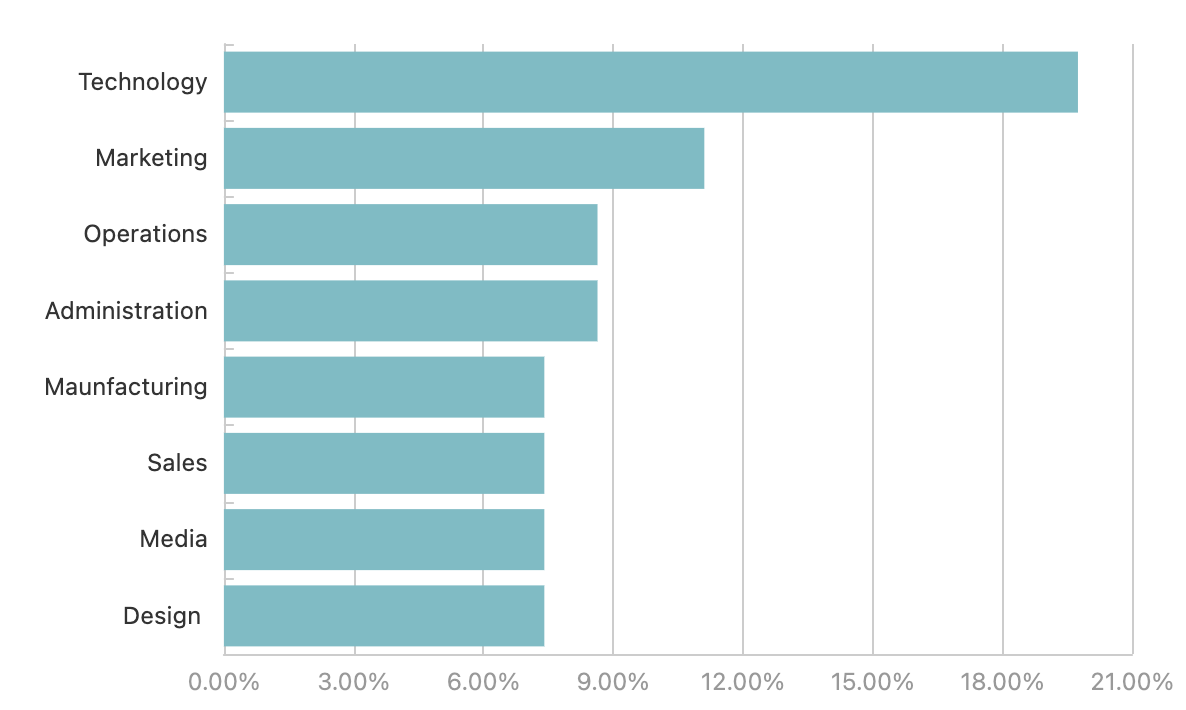}
    \caption{Distribution of occupations with high proficiency requirements for ChatGPT-related skills}
    \label{fig:3}
\end{figure}

The analysis shows that certain industries and occupations will have high proficiency requirements for ChatGPT-related skills. Occupations such as advertising, interactive design, game numerical planning, full-stack engineering, community operations, artificial intelligence, and legal specialist will require candidates to be proficient in ChatGPT and other AIGC models. Figure~\ref{fig:3} shows the primary industry distribution of these occupations, including technology, marketing, operations, administration, and design industries. For example, advertising positions will increasingly require candidates to use AI tools like ChatGPT and Midjourney to complete tasks such as topic planning, copywriting, visual design, and video production, to improve work efficiency.

\subsubsection{Occupations with low proficiency requirements for ChatGPT-related skills}\label{subsubsec3}

\begin{figure}[!h]
    \centering
    \includegraphics[width=10cm]{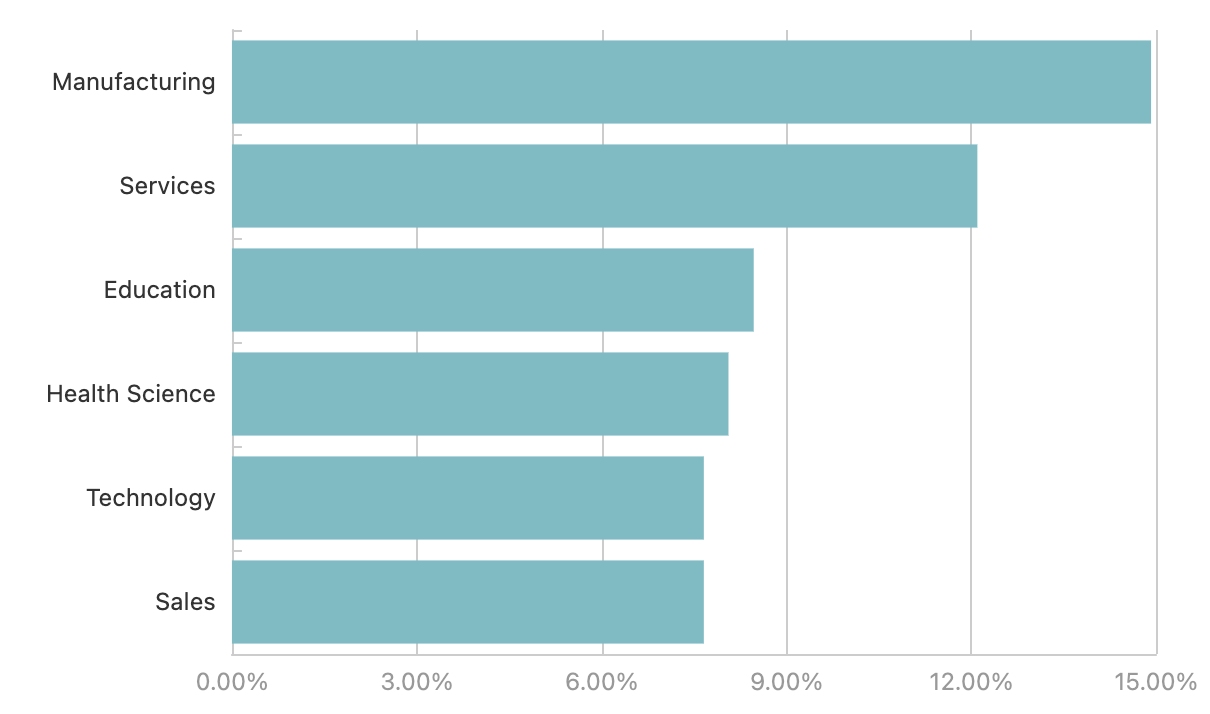}
    \caption{Distribution of occupations with low proficiency requirements for ChatGPT-related skiils}
    \label{fig:4}
\end{figure}

While some occupations may have low proficiency requirements for ChatGPT-related skills, the potential for ChatGPT's application and development in these industries should not be underestimated. Even though job seekers in fields such as manufacturing, services, education, and health science may not currently be required to use ChatGPT (Figure~\ref{fig:4}), it is possible that they may benefit from its use in the future. For example, welders and drillers may use ChatGPT to access technical manuals and guides, while restaurant staff and lifeguards could utilize ChatGPT for language translation in order to better communicate with customers. Additionally, performing arts teachers, dental consultants, plastic surgeons, and ophthalmologists may use ChatGPT for research and learning purposes to improve their practices. Therefore, having a basic understanding of ChatGPT and its potential uses could be beneficial for job seekers in a wide range of industries.

\subsubsection{Salary of occupations enabled by ChatGPT}\label{subsubsec4}

There is a significant positive correlation between the level of proficiency required for ChatGPT in the future and the average monthly salary of occupations. Occupations with higher salary are more likely to require higher level of ChatGPT proficiency in the future. For example, the average monthly salary of interactive designers, game data planners, full-stack engineers, and community operators, positions that will require a high level of ChatGPT proficiency, all exceed 13,000 RMB.

\section{Discussion}\label{sec3}

On the basis of the prediction results, it is evident that ChatGPT is a promising AI technology with great potential for application in the workplace. Its adoption can bring significant benefits to organizations, including improved work efficiency, increased creativity and innovation, and increased professional abilities and competitiveness for employees. As a result, it is likely that ChatGPT will become a basic skill requirement for jobs related to technology, marketing, operations, and design. Job seekers who are proficient in using ChatGPT will have a competitive advantage in the labor market. However, improving ChatGPT skills requires a dedicated effort that involves practicing, reading user manuals, learning the basics of Natural Language Processing (NLP), and programming skills. Thus, individuals who aim to develop ChatGPT-related skills must be willing to invest in continuous learning and improvement.

As ChatGPT becomes increasingly integrated into the workplace, it has the potential to transform the way we work and interact with AI. However, like with the introduction of new tools throughout history, we do not need to fear that ChatGPT will replace our work entirely. Instead, the development of AI technology will bring new job opportunities and change the nature of existing jobs. It is important for workers to embrace these changes and actively learn new technologies and skills to stay competitive in the future workplace. By doing so, we can seize the development opportunities that ChatGPT and other emerging AI technologies bring to the table.

\section{Limitations}\label{sec4}

It is important to recognize that our research has certain limitations that need to be considered. The primary constraint is the relatively short timeframe during which we collected the data, given that ChatGPT has been available for less than six months and there are insufficient job postings for a fully comprehensive analysis. Nonetheless, our research provides valuable insight into the expected labor market patterns and the ChatGPT-related skills that may be required in the future. As more job postings become available, our research can be refined and expanded to provide a more accurate and complete analysis. Despite the limitations, we believe that our research offers a solid foundation for further exploration of this topic, highlighting the need for ongoing research and analysis to better understand the impact of AI technologies like ChatGPT on the labor market.

\section{Methods}\label{sec5}

\subsection{Problem Formulation}
According to the occupation knowledge graph developed by BOSS Zhipin Career Science Lab (CSL), the occupation-skill incidence matrix can be defined as:
\begin{equation}
    W_{ij} = \begin{cases}
   r_{ij} &\text{if } S_j \to O_i \\
   0 &\text{if } S_j \not \to O_i
\end{cases}
, i \in (1,l_o), j \in (1,l_s),
\end{equation}
where $O_i$ is the i-th occupation, $S_j$ is the j-th skill, $l_o$ is the number of occupations, $l_s$ is the number of skills, $S_j \to O_i$ means $S_j$ is related to $O_i$ and $r_{ij}$ is the degree of demand between $S_j$ and $O_i$ in the given knowledge graph.

Our goal is to predict the degree of demand for each skill for all occupations using the existing incidence matrix. Specifically, we aim to predict the $W_{ij}$ value where $W_{ij} = 0$, in order to complete the incidence matrix $W$.

\subsection{Problem Analysis}
To predict the degree of demand, we can utilize the collaborative filtering algorithm, such as matrix decomposition, to extract hidden features and predict the weight of unknown edges in the knowledge graph.
However, direct matrix decomposition using singular value decomposition (SVD) faces a challenge when there is insufficient data. In such cases, the original matrix may not provide enough information to accurately fill in unmarked data. Another challenge arises when the number of predicted occupations significantly exceeds the number of observed occupations. 

FunkSVD can improve the computational efficiency and data sparsity issues of SVD. It decomposes the incidence matrix into two low-dimensional matrices. The reconstructed low-dimensional matrix predicts the degree of occupation demand for skills. For the sample data is imbalanced, it becomes necessary to include semantic features of occupations and skills to guide the matrix decomposition process. By integrating these features, we can better understand the complex interdependencies between occupations and skills, leading to more precise predictions of the demand for ChatGPT in a wide range of occupations.

\subsection{FunkSVD Algorithm Fused with Semantic Information}
We adopt the idea of funkSVD\cite{funkSVD} to obtain the feature matrices of occupations and skills, respectively. 
\begin{equation}
    H_{o_i}H_{s_j}^T = r_{ij},
\end{equation}
where $H_{o_i}$ and $H_{s_j}$ denotes the hidden features of $O_i$ and $S_j$ respectively. Feature matrices essentially summarize information about occupations or skills, with each dimension of the matrix representing a different feature. Using the feature matrices, we further complete the occupation-skill incidence matrix and obtain the demand level of all skills for each occupation. 
 
 However, due to the sparsity of the incidence matrix and the fact that the amount of data to be predicted is several times that of the observed data, we utilize the embedding layer of the BERT\cite{devlin-etal-2019-bert} and TextCNN\cite{kim-2014-convolutional} to extract the semantic features $T_s$ and $T_o$
\begin{equation}
    E_{o_i} = \rm{Embedding}(\it O_i), E_{s_i} = \rm{Embedding}(\it S_i),
\end{equation}
\begin{equation}
    T_{o_i} = \rm maxpool(conv(\it E_{o_i})), T_{s_i} = \rm maxpool(conv(\it E_{s_i})),
\end{equation}
where $\rm conv$ is a convolution layer and $\rm maxpool$ is a max pooling layer.
These features are then incorporated into the feature matrices, increasing the dimensions of the feature matrices and enabling a more accurate prediction of the degree of occupations' demand for skills:
\begin{equation}
    r_{ij} = [H_{o_i}:T_{o_i}][H_{s_j}:T_{s_j}]^T,
\end{equation}
where $[\cdot:\cdot]$ denotes the concatenation operation.
By integrating semantic information with matrix decomposition techniques, we can more effectively predict the demand for ChatGPT in a wide range of occupations.

\subsection{Objective}
Based on the idea of funkSVD, the correlations can be predicted by the feature matrices $P=[H_s:T_s]$ and $Q=[H_o:T_o]$.
\begin{equation}
    \hat{W}=PQ^T,
\end{equation}
Afterward, we define the loss function as the square error between the predicted matrix $\hat{W}$ and the true matrix $W$, using the labeled nodes as reference points.
\begin{equation}
    \mathcal{L}=\sum_{W_{ij} \not = 0}{||\hat{W}_{ij}-W_{ij}||^2}.
\end{equation}
Finally, the parameter matrix is updated using stochastic gradient descent.

\bmhead{Acknowledgments}

We would like to express our gratitude to our colleagues at BOSS Zhipin Career Science Lab (CSL) for their invaluable insights and discussions.

%%===========================================================================================%%
%% If you are submitting to one of the Nature Portfolio journals, using the eJP submission   %%
%% system, please include the references within the manuscript file itself. You may do this  %%
%% by copying the reference list from your .bbl file, paste it into the main manuscript .tex %%
%% file, and delete the associated \verb+\bibliography+ commands.                            %%
%%===========================================================================================%%

\bibliography{sn-article}% common bib file
%% if required, the content of .bbl file can be included here once bbl is generated
%%\input sn-article.bbl

\end{document}